\documentclass[epj]{svjour}

\usepackage{latexsym}
\usepackage{graphics}
\usepackage{cite}
\usepackage{amsmath}

\begin{document}

\title{A recoil detector for the measurement of antiproton-proton elastic scattering at angles close to 90$^{\circ}$}

\author{Q. Hu\inst{1,2,6}, U. Bechstedt\inst{2}, A. Gillitzer\inst{2}, D. Grzonka\inst{2}, A. Khoukaz\inst{5}, F. Klehr\inst{4}, A. Lehrach\inst{2}, D. Prasuhn\inst{2},  J. Ritman\inst{2,3}, T. Sefzick\inst{2}, T. Stockmanns\inst{2}, A. T\"aschner\inst{5}, P. Wuestner\inst{4},  H. Xu\inst{2,}\thanks{\emph{Corresponding author: h.xu@fz-juelich.de}}  
}                     

\institute{ Institute of Modern Physics, Chinese Academy of Sciences, Lanzhou, 730000, China \and 
		Institute f\"ur Kernphysik, Forschungszentrum J\"ulich, J\"ulich, 52425, Germany \and
		Ruhr-Universit\"at Bochum, Bochum, 44780, Germany \and
		Zentralinstitut f\"ur Engineering, Elektronik und Analytik, Forschungszentrum J\"ulich, J\"ulich, 52425, Germany \and
		Institut f\"ur Kernphysik, Universit\"at M\"unster, M\"unster, 48149, Germany \and
		University of Chinese Academy of Sciences, Beijing, 100049, China
		 							}
\date{Received: date / Revised version: date}

\abstract{
The design and construction of a recoil detector for the measurement of recoil protons of  antiproton-proton elastic scattering at scattering angles close to 90$^{\circ}$ are described. The performance of the recoil detector has been tested in the laboratory with radioactive sources and at COSY with proton beams by measuring proton-proton elastic scattering. The results of laboratory tests and commissioning with beam are presented. Excellent energy resolution and proper working performance of the recoil detector validate the conceptual design of the KOALA experiment at HESR to provide the cross section data needed to achieve a precise luminosity determination at the PANDA experiment.
}

\authorrunning{Q. Hu etc.}
\titlerunning{A recoil detector for the measurement of the antiproton-proton elastic scattering at angles close to 90$^{\circ}$}
\maketitle

\section{Introduction}
\label{intro}

The High-Energy Storage Ring (HESR) at the Facility for Antiproton and Ion Research (FAIR) will deliver high quality antiproton beams over a momentum range from 1.5 to 15 GeV/c~\cite{panda}. The PANDA experiment at HESR will perform high precision investigations of antiproton annihilation on internal proton and nuclear targets to pursue various topics covering the weak and strong forces, exotic states of matter and the structure of hadrons~\cite{panda}. 

In order to serve the wide physics potential with antiprotons at HESR, PANDA is designed as a general purpose detector covering nearly the complete solid angle for both charged and neutral particles with good momentum resolution, particle identification capability as well as excellent vertex determination. In order to determine the integrated luminosity with a higher precision than $5\%$ achievable with Schottky measurements~\cite{schottky}, a concept for a luminosity monitor based on measuring elastic scattering in the Coulomb-nuclear interference region has been presented~\cite{lumi1}. This system is based on tracking the antiprotons scattered near the beam axis with a multiple layer setup based on high-voltage monolithic active pixel sensors (HV-MAPS)~\cite{lumi2}. The detector will be located about 10~m downstream from the interaction point and cover the polar angle range $\theta$=3--8 mrad. Based on two body kinematics the squared 4-momentum transfer $t$ can be determined  from the measured polar angle of the elastically scattered antiprotons. Simulation studies  indicated that the precision of the luminosity determination for the concept is limited by the small acceptance range of the squared 4-momentum transfer $t$ covered by the detector. In order to achieve the desired absolute luminosity precision $\frac{\Delta L}{L}\le3\%$ requested by physics goals~\cite{panda}, it requires the knowledge of the physics quantities $\sigma_{\text{tot}}$, $\rho$ and $b$ describing the dependence of the antiproton-proton elastic cross section on $t$ in the Coulomb-nuclear interference region. To alleviate the lack of existing data on antiproton-proton interactions in the momentum region of PANDA, an independent experiment called KOALA ($\bf{K}$ey experiment f$\bf{O}$r P$\bf{A}$NDA $\bf{L}$uminosity determin$\bf{A}$tion) at HESR, dedicated to the measurement of antiproton-proton elastic scattering, has been proposed~\cite{day1}. 

The KOALA experiment at HESR will measure a large range of the squared 4-momentum transfer $t$-distribution of antiproton-proton elastic scattering in order to determine the parameters of total cross section $\sigma_{\text{tot}}$, the ratio, $\rho$,  of the real part to the imaginary part of the forward elastic scattering amplitude and nuclear slope $b$ (see sect.~\ref{principle}). The idea is to measure the scattered beam particles at forward angles by tracking detectors and the recoil target protons near 90$^{\circ}$ by energy detectors. The PANDA luminosity monitor detector can be used for the forward measurement and the recoil protons will be measured by a new recoil detector. The recoil detector will measure both the kinetic energy and the polar angle of the recoil protons to achieve a stronger background suppression since $t$ is directly proportional to the proton's kinetic energy $T_{\text{p}}$, i.e. $\left|t\right|=2m_{\text{p}}T_{\text{p}}$. 

In order to test the proposed method one of the recoil arms of the KOALA experiment at HESR has been designed and built. It was commissioned at COSY~\cite{cosy} by measuring proton-proton elastic scattering since the beam momenta are similar and the recoil particles are the same for antiproton-proton elastic scattering at HESR and for proton-proton elastic scattering at COSY. 
 
\section{Concept and design of the recoil detector}
\label{concept}
\subsection{Principle of the concept}
\label{principle}

One classical way to determine the absolute luminosity is to find a reference channel which has a well known absolute cross section, for example, Bhabha scattering~\cite{bhabha} for e$^{+}$e$^{-}$ annihilation experiments. However, there is no ideal reaction for hadronic reactions.  Another traditional method to determine the absolute luminosity is to measure the rate of elastic scattering. This is related to the total interaction rate by the optical theorem, which states that the total cross section is directly proportional to the imaginary part of the forward elastic scattering amplitude extrapolated to zero momentum transfer: $\sigma_{\text{tot}}=4\pi \cdot $Im$[(f_{\text{el}}(t=0)] $. This idea, for instance, was pursued by the TOTEM experiment at LHC with an achieved precision for the absolute luminosity determination of 4$\%$~\cite{totem1, totem2}. 

The method of using the optical theorem requires measuring not only the elastic scattering at small polar angles but also the total interaction rate with high precision. This method might not give the highest possible precision for the luminosity determination for the PANDA experiment due to the limited pseudorapidity coverage, i.e. the maximum acceptance of 1$^{\circ}<\theta<140^{\circ}$ of the PANDA tracking detectors~\cite{MVDTDR,STTTDR}. Therefore, an alternative method the so-called Coulomb-normalization is suggested. The Coulomb cross section of antiproton-proton elastic scattering is sufficiently accurately known. However, Coulomb scattering is always accompanied by the strong interaction, which is poorly known. By measuring the antiproton-proton elastic scattering rate down to the Coulomb dominated region, e.g. more than 98$\%$ of the cross section from Coulomb scattering, the absolute luminosity can be normalized with a precision of 2--3$\%$ with the following relation between the measured differential elastic scattering rate $\frac{\text{d}R_{\text{el}}}{\text{d}t}$ and luminosity ${L}$, 
\begin{equation}
\frac{\text{d}R_{\text{el}}}{\text{d}t}=\frac{\text{d}\sigma_{\text{el}}}{\text{d}t}{L}.
\end{equation}

\begin{figure}
\begin{center}
\resizebox{0.48\textwidth}{!}{
    \includegraphics{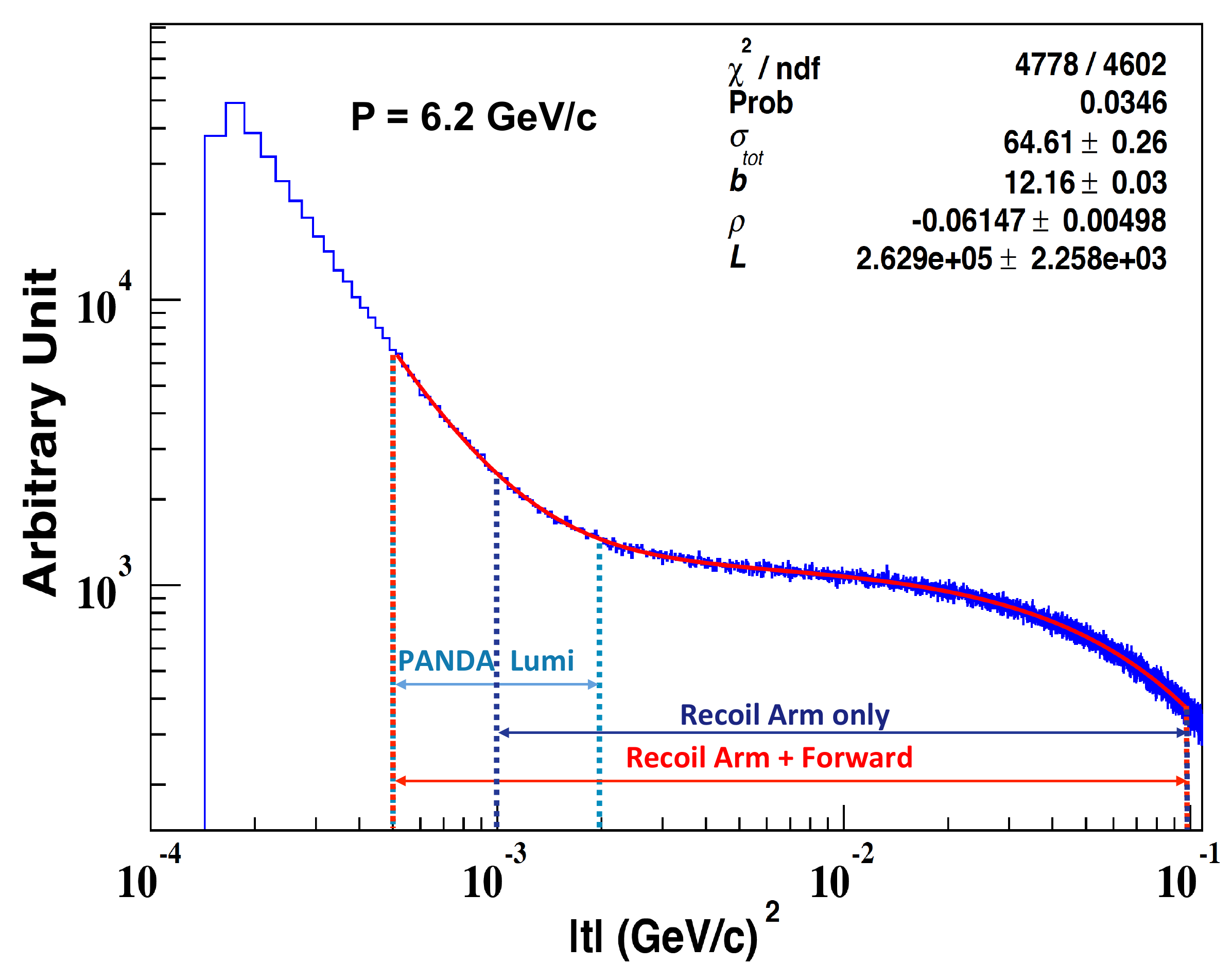}
}
\end{center}
\caption{A simulated $t$-distribution of antiproton-proton elastic scattering at 6.2 GeV/c beam momentum. The measurable $t$ ranges by the PANDA luminosity monitor detector (light blue) and by the KOALA experiment (red) are indicated.}
\label{fig:1}       
\end{figure}

Technically it is extremely difficult to measure the elastic events in such small $t$ in which the Coulomb scattering dominates, i.e. $\left|t\right|<$0.0002~(GeV/c)$^{2}$ or $\theta<$0.15$^{\circ}$, at the beam momenta relevant for the HESR. Instead of pursuing pure Coulomb scattering it is still possible to determine the luminosity by analysing the characteristic shape of the $t$-distribution in a large range, e.g. 0.0005$<$$\left|t\right|$$<$0.1~(GeV/c)$^{2}$. The $\frac{\text{d}R_{\text{el}}}{\text{d}t}$ distribution will be measured by experiment. The antiproton-proton elastic scattering with conventional parameterization can be described in terms of the Coulomb and nuclear amplitudes, $f_{\text{c}}$ and $f_{\text{n}}$, by means of dispersion relations. At small $t$ one obtains~\cite{Block, E760}
\begin{equation}
 \frac{\text{d}\sigma}{\text{d}t}=\frac{\pi}{k^2}\left|f_{\text{c}}e^{i\delta}+f_{\text{n}}\right|^2 = \frac{\text{d}\sigma_{\text{c}}}{\text{d}t}+\frac{\text{d}\sigma_{\text{int}}}{\text{d}t}+\frac{\text{d}\sigma_{\text{n}}}{\text{d}t},
\end{equation}
where 
\begin{equation}
\frac{\text{d}\sigma_{\text{c}}}{\text{d}t}=\frac{4\pi\alpha_{\text{EM}}^2G^4(t)(\hbar c)^2}{\beta^2t^2},
\end{equation}

\begin{equation}
\frac{\text{d}\sigma_{\text{int}}}{\text{d}t}=\frac{\alpha_{\text{EM}}\sigma_{\text{tot}}}{\beta\left|t\right|}G^2(t){\text{e}}^{-\frac{1}{2}b\left|t\right|}(\rho\cos\delta+\sin\delta),
\end{equation}
and
\begin{equation}
\frac{\text{d}\sigma_{\text{n}}}{\text{d}t}=\frac{\sigma_{\text{tot}}^2 (1+\rho^2){\text{e}}^{-b\left|t\right|}}{16\pi(\hbar c)^2}.
\end{equation}

Here, $\alpha_{\text{EM}}$ is the fine structure constant. $G(t)$ is the proton dipole form factor defined as $G(t)=(1+\Delta)^{-2}$, with $\Delta\equiv\left|t\right|/0.71$ (GeV/c)$^2$. $\delta (t)$ is the Coulomb phase~\cite{Cahn},
\begin{equation}
\delta(t)=\alpha_{\text{EM}}\left[0.577+\ln\left(\frac{b\left|t\right|}{2}+4\Delta\right)+4\Delta\ln(4\Delta)+2\Delta\right].
\end{equation}
The Coulomb cross section is a known function of $t$. The parameters $\sigma_{\text{tot}}$, $\rho$ and $b$ involved in the parameterisation of the nuclear and the interference cross section need to be determined by experiment.

\begin{figure*}
\begin{center}
\resizebox{0.97\textwidth}{!}{
\includegraphics{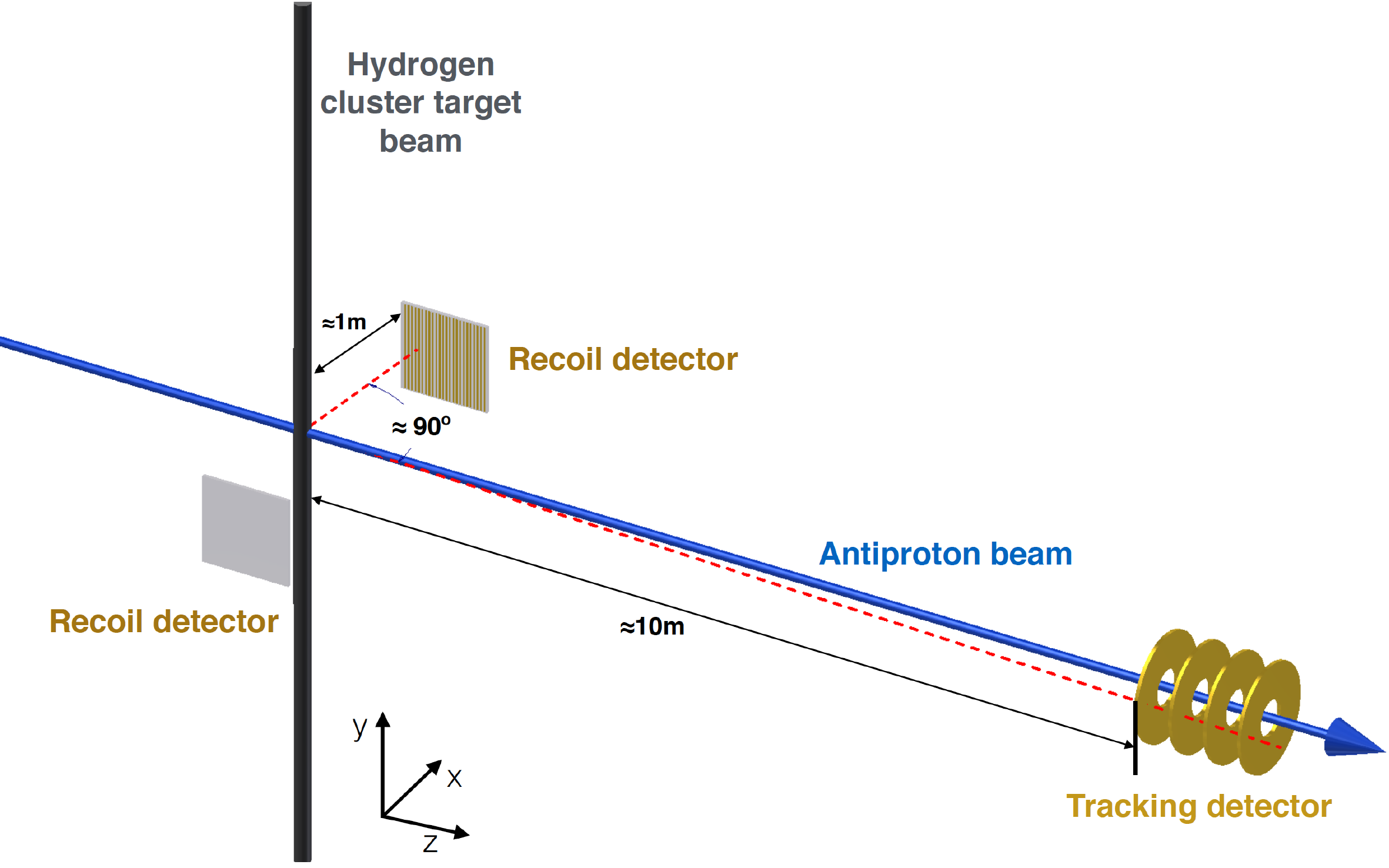}
}
\end{center}
\caption{Sketch of the KOALA experiment at HESR.}
\label{fig:2}    
\end{figure*}

A simulated $t$-distribution of antiproton-proton elastic scattering with 6.2~GeV/c beam momentum is shown in Fig.~1. This has been fit in the range of $\left|t\right|$=0.0005--0.1~(GeV/c)$^{2}$ by using the formula above. Not only the luminosity, but also the parameters $\sigma_{\text{tot}}$, $\rho$ and $b$ are determined with a precision better than 1\%. However, limited by the detector acceptance, the PANDA luminosity monitor can only measure a small range of the $t$-distribution, as noted by the light blue dashed lines in Fig.~1. Due to strong correlations between the fit parameters it is difficult to determine the luminosity together with the unknown parameters within such a small range of the $t$-distribution. In order to achieve the desired precision of 3\% for the luminosity determination it is necessary to measure the 3 parameters. Therefore, the KOALA experiment at HESR has been proposed to determine the unknown parameters by measuring a large range of the $t$-distribution. The $t$-range to be measured by the KOALA experiment is also depicted by the red dashed lines in Fig.~1. The desired range of the $t$-distribution will be measured by combining the measurements of the recoil arm and forward detector.

A sketch of the KOALA experiment at HESR is shown in Fig.~2. The forward measurement will track the scattered antiprotons, just as the PANDA luminosity monitor. Therefore, the PANDA luminosity monitor detector~\cite{lumi2} could be employed as the forward arm of the KOALA experiment. In addition, a recoil arm to detect the recoil protons is needed. In contrast to the forward measurement, the recoil detector will measure the angle and the energy of the recoil protons  within an angular range (recoil angle defined as $\alpha \equiv 90^{\circ}-\theta_{\text{lab}}$) $0^{\circ}<\alpha<19^{\circ}$.

\subsection{Design of the recoil detector}
\label{sec:3}

The measurement goals of the KOALA experiment at HESR dictate the following design criteria for the recoil detector.
\begin{itemize}
  \item The recoil detector should cover the full range of the expected recoil angles up to $\alpha$=$19^{\circ}$.
  \item The recoil detector should be sensitive to the full interaction volume. The interaction volume consists of the intersection of a cluster jet target of 9 mm transverse width~\cite{Khoukaz} and 1 mm longitudinal thickness~\cite{TargetTDR} with a perpendicular antiproton beam of about 5 mm diameter. We refer to the centroid of the interaction volume as the interaction "point".
  \item The individual elements of the recoil detector should subtend only a small fraction of recoil angles, $\Delta \alpha < 0.1^{\circ}$, from the interaction "point" because of the rapid variation of $\frac{\text{d}\sigma_{\text{el}}}{\text{d}t}$ in the small $t$ region.
  \item The energy resolution of the recoil detector should be better than 30~keV due to the fine granularity of detectors along the beam axis.
  \item The recoil detector should be able to measure the recoil protons with energy down to about 600~keV, which corresponds to $\left|t\right|$=0.001~(GeV/c)$^{2}$.
  \end {itemize} 

\begin{figure}
\begin{center}
\resizebox{0.48\textwidth}{!}{
  \includegraphics{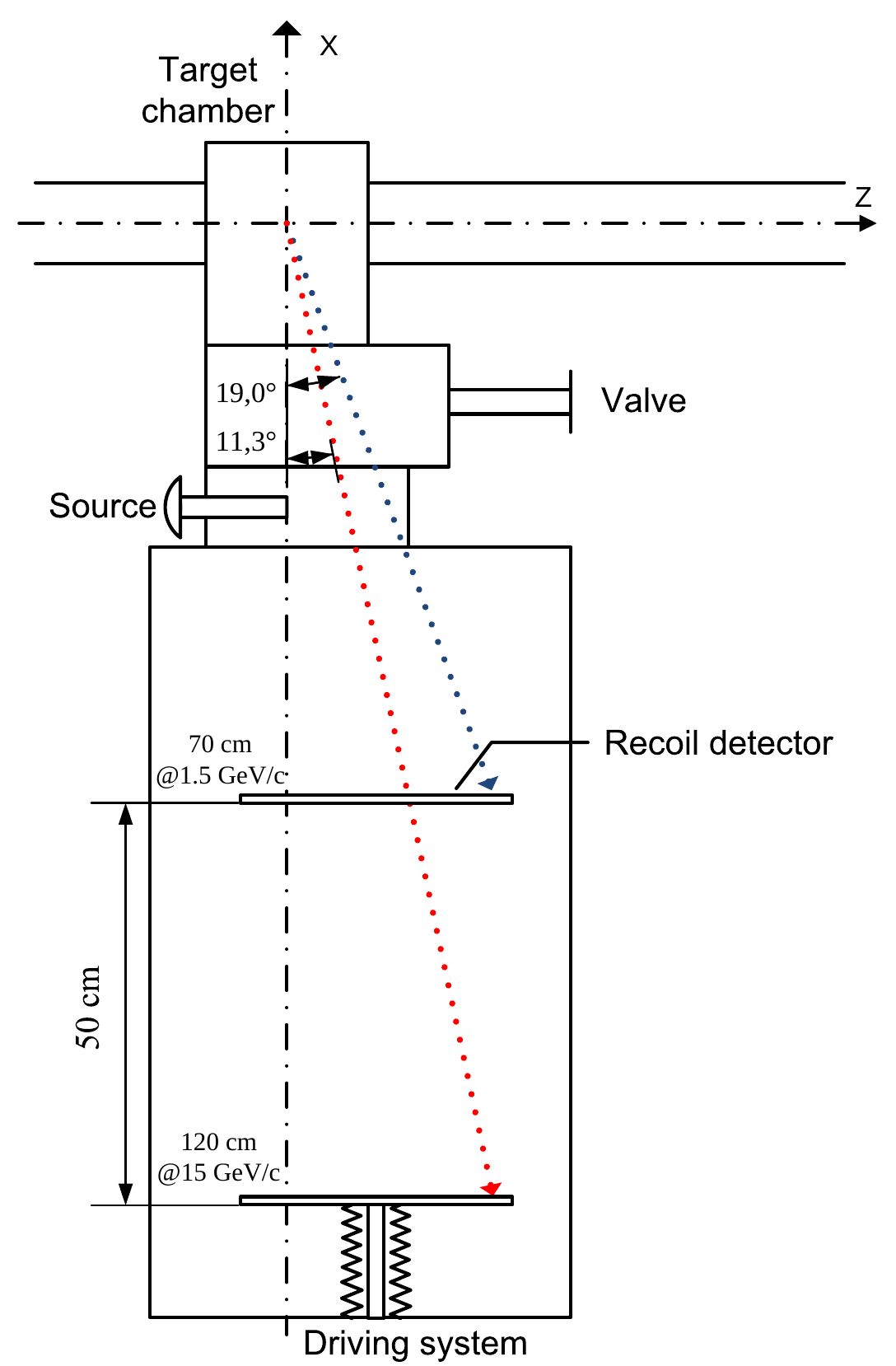}
}
\end{center}
\caption{Schematic view of the recoil detector showing the complete setup which has a movable detector plane in order to cover the desired range of the recoil angle depending upon the chosen beam momentum.}
\label{fig:3}      
\end{figure}

\begin{figure}
\begin{center}
\resizebox{0.5\textwidth}{!}{
  \includegraphics{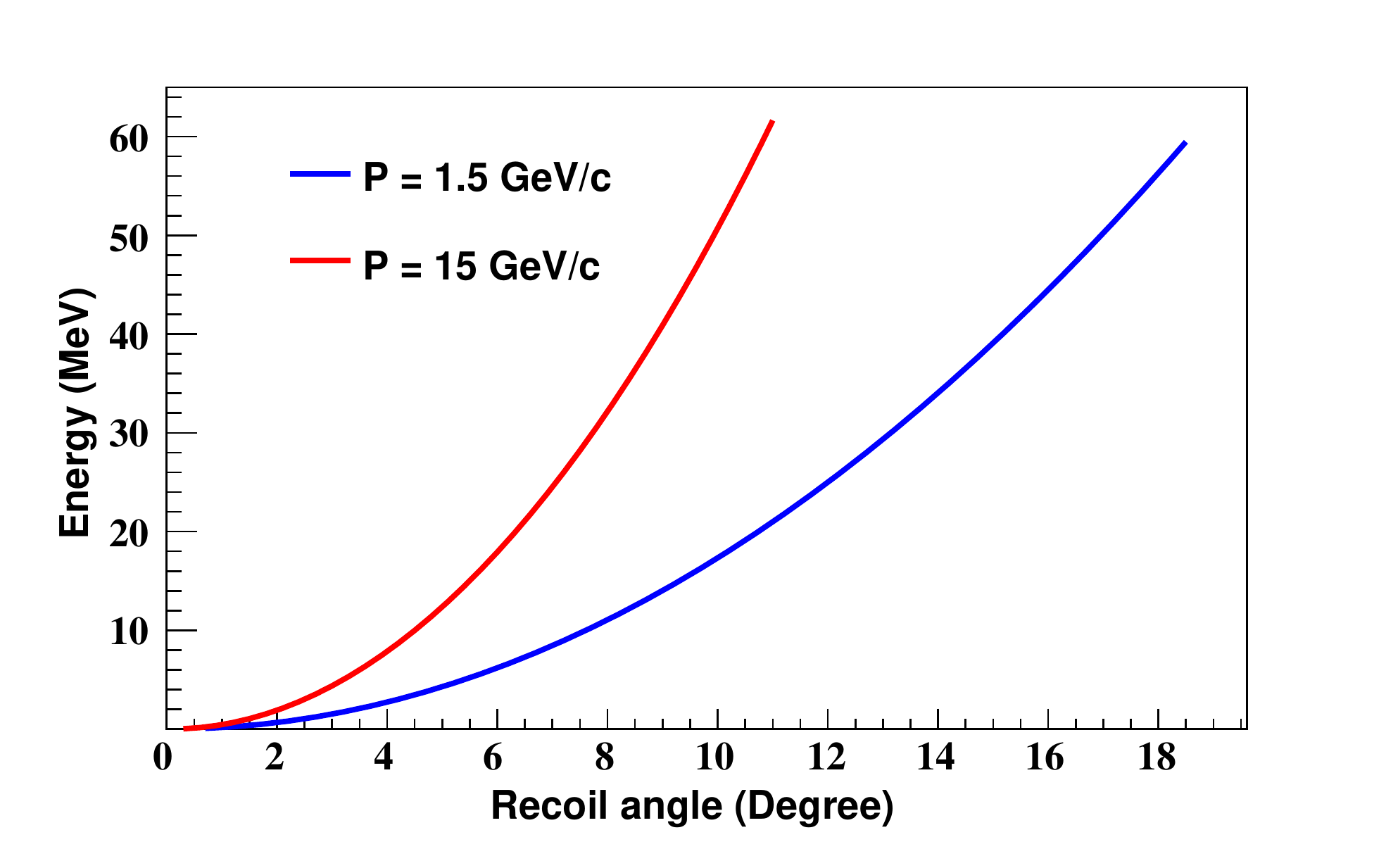}
}
\end{center}
\caption{The kinetic energy of the recoil protons as a function of the recoil angle at beam momenta P=1.5 and 15~GeV/c, blue and red, respectively.}
\label{fig:4}       
\end{figure}

\begin{figure}
\begin{center}
\resizebox{0.48\textwidth}{!}{
  \includegraphics{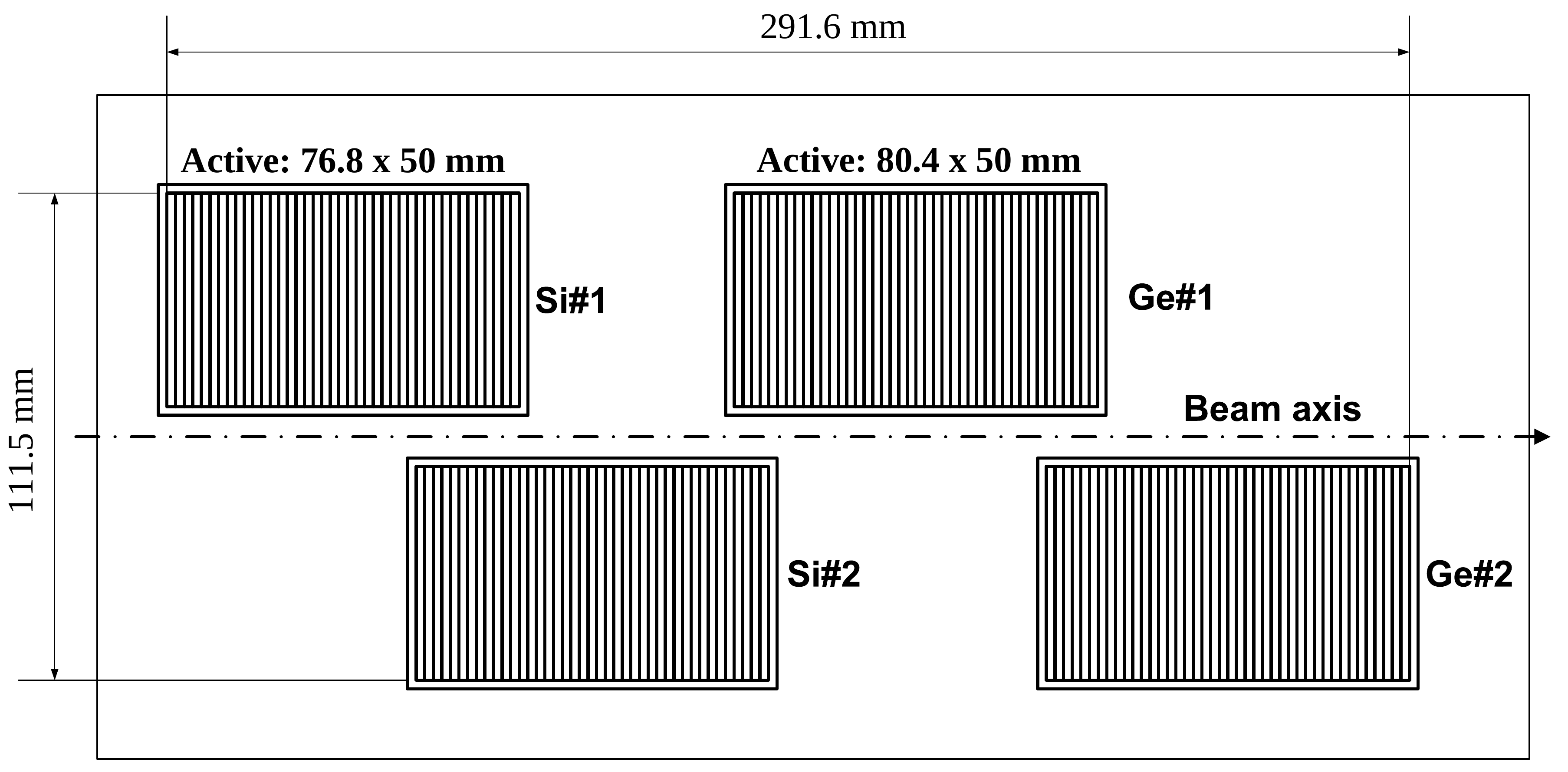}
}
\end{center}
\caption{The proposed detector layout (view in the YZ plane) consisting of 2 silicon strip detectors with 1 mm thickness and 2 germanium strip detectors with 5 and 11 mm thickness.}
\label{fig:5}     
\end{figure}

Considering all these requirements, the recoil detector assembly has been designed as presented in Fig.~3. With a movable plate the detector can cover various ranges of recoil angle as required for the beam momenta being investigated. The kinetic energy of recoil protons as a function of the recoil angle at beam momenta P=1.5 and 15~GeV/c is plotted in Fig.~4. In order to detect the recoil protons with kinetic energy in the range of 0.6--60~MeV, corresponding to $\left|t\right|$ = 0.001--0.1~(GeV/c)$^{2}$, 2 silicon strip detectors with 1 mm thickness and 1.2 mm pitch and 2 germanium strip detectors with 5 and 11 mm thickness and 1.2 mm pitch have been proposed. Thus, not only the energy but also the relevant angular information of the recoil protons will be obtained. As shown in Fig.~5, the silicon and germanium detectors are arranged in two overlapping rows. It is noted that the different size of the overlapping areas between the two rows is because of the limited thickness of the silicon detectors. The 1 mm thick silicon detectors will measure the recoil protons with energy up to 12~MeV. The protons with higher energy up to 60~MeV will be measured by 5 and 11 mm thick germanium detectors. Those detectors will be positioned in a dedicated vacuum chamber and can be placed at a distance of 70 cm to 120 cm from the interaction point, depending upon the momentum of the HESR beam.

\section{Construction of the recoil detector}
\label{sec:construction}

\subsection{Energy detector}
\label{sec:detector}

A picture of the recoil detector as used for the commissioning is shown in Fig.~6. Two silicon strip sensors  produced by MICRON~\cite{micron} with customized dimensions of 76.8 mm (length) x 50 mm (width) x 1 mm (thickness) have been placed at about 1 m from the target to cover the recoil angles, ${\alpha}$=0$^{\circ}$--5.7$^{\circ}$. This corresponds to the region in which the silicon detector can stop the recoil protons with energies up to 12~MeV. Each silicon detector has 64 strips with 1.2 mm pitch. The silicon sensors were glued on ceramic frames which have been screwed to the aluminium detector holder. In addition, two germanium strip detectors produced by SEMIKON~\cite{semikon} with 5 and 11 mm thickness have been set up in 2 rows as well. They can measure the recoil protons with energies up to 60~MeV. Both germanium detectors have a strip width of 1.2 mm and offer 67 readout strips in a sensitive area of 80.4 mm (length) x 50 mm (width). Each germanium detector has been fabricated on a gold plated copper frame, which enables excellent thermal conductivity.

\begin{figure}
\begin{center}
\resizebox{0.48\textwidth}{!}{
  \includegraphics{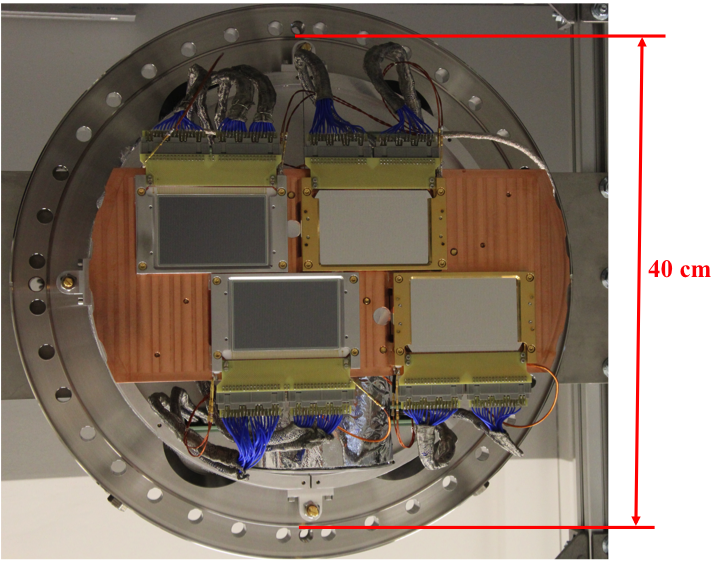}
}
\end{center}
\caption{Recoil detector assembled for a beam test at COSY. The two germanium detector with gold plated copper frames are fixed at the right side of the copper cold plate and two silicon detectors are located at the left side.}
\label{fig:6}      
\end{figure}

The detector strips have been wire-bonded to an adaptor PCB board for signal readout. A standard 3M four wall header with 34 pins has been employed as a socket on the PCB board for the further readout cabling. All detectors have been fixed to the copper cold plate with indium foil to ensure good thermal conduction. Each detector has a PT100 sensor integrated in its holder to monitor the operating temperature.

\subsection{Detector chamber}
\label{sec:chamber}
 
The detector chamber added to the existing target chamber~\cite{anke} for the tests with COSY beam is drawn in Fig.~7. Due to space limitations and the restricted opening angle of the target chamber, the detector is held at a fixed location for the commissioning. The detector plate is located horizontally at a distance of 1 m from the interaction point. With an optimal design all detectors as well as the cold plate are integrated on one main flange as presented in Fig.~6. The detector chamber is typically operated at the vacuum pressure of $10^{-8}$~mbar for beam tests. One gate valve has been installed between the recoil detector chamber and the target chamber to enable maintenance on the recoil detector chamber without breaking the COSY vacuum. A linear motion feed-through rod has been added to the chamber to hold an alpha radioactive source for calibrating and monitoring the detectors. 

\begin{figure}
\begin{center}
\resizebox{0.48\textwidth}{!}{
  \includegraphics{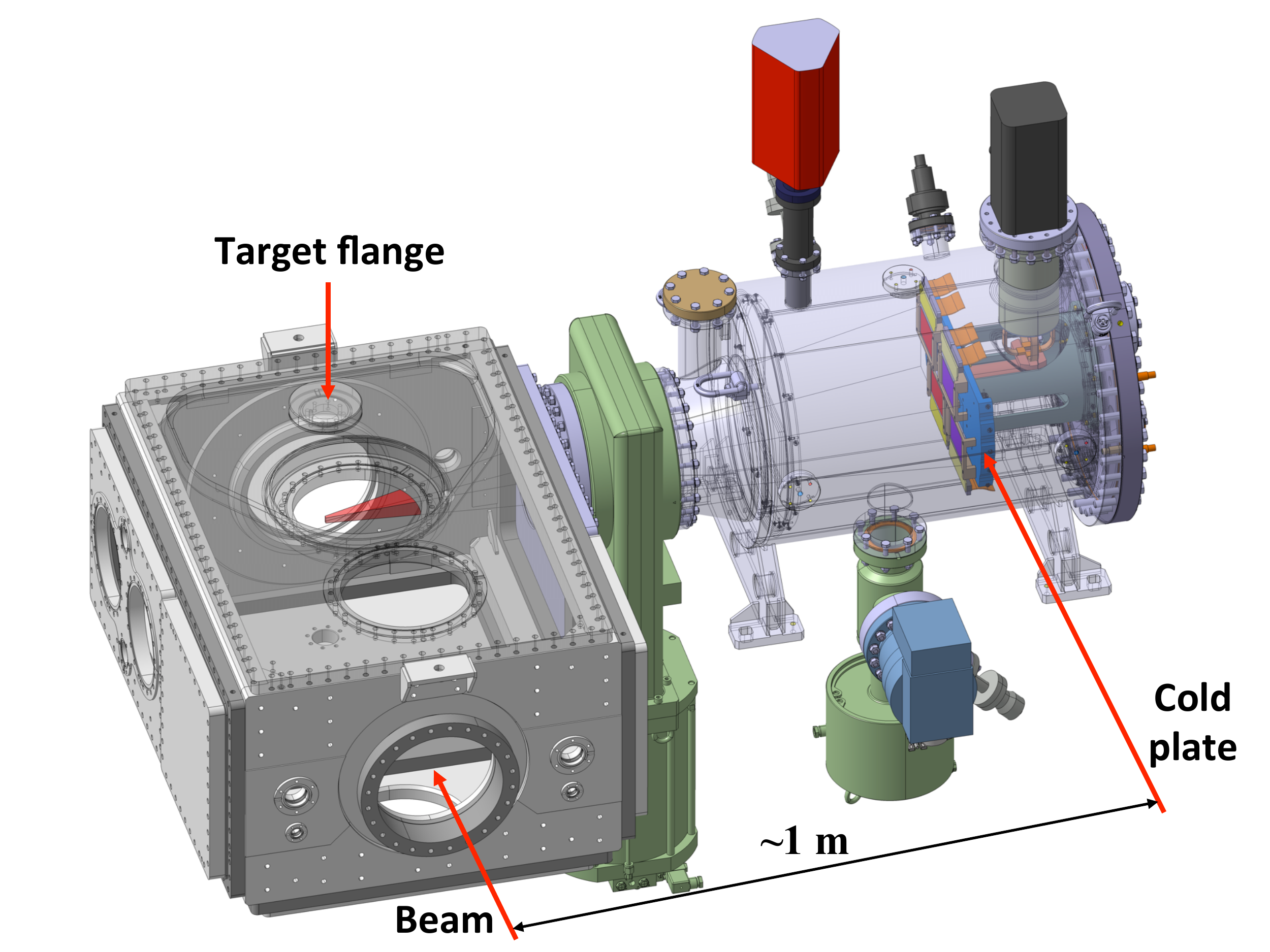}
}
\end{center}
\caption{Detector chamber (right) designed to match the existing target chamber (left).}
\label{fig:7}   
\end{figure}

\subsection{Temperature control}
\label{sec:temperature}

A pulse tube PT30 Cryorefrigerator from CRYOMECH~\cite{cryomech} with maximum cooling power of 30~W at 70~K has been employed to cool the detector system. The cold plate of the detectors has been coupled by flexible copper bands to the cold head in order to reduce vibrations from the cold head. Tests of the PT30 Cryorefrigerator showed that it took about 25 minutes to reach 50~K without any load. Further integration tests have proven that there was no significant influence from vibration of the cold head on the detector performance.

A high precision Lake Shore Model 336 temperature controller~\cite{lakeshore} has been chosen for the temperature measurement and control. It provides a maximum heating power of 100~W on channel 1 and a 50~W heating output of channel 2. Two 50~ohm heating resistors were mounted on the cold plate to adjust the temperature of the recoil detector. The reference temperature for the LSM-336 temperature controller was provided by two PT100 sensors fixed on the cold plate. Another four PT100 sensors were integrated on the detector holders to be monitored by the LSM-336 controller. The temperature differences between the PT100 sensors on the cold plate and on the germanium detector holders were less than 0.6~K, which is consistent with the feature of a PT100 sensor. The temperature controller was remotely accessed and controlled via internet. With the combination of the PT30 Cryorefrigerator and the LSM-336 temperature controller the working temperature of the recoil detector could be set to any desired value in the range of 70--300~K. The excellent performance of the temperature control system enabled convenient operation of the recoil detector system during the experiment. 

\subsection{FEE and DAQ}
\label{sec:FEE}

Electronics modules provided by MESYTEC~\cite{mesytec} have been selected for the signal readout of  the silicon and germanium strip detectors. To read out all strips and the rear sides of the detectors, a total readout of 180 signal channels has been implemented for the commissioning experiment. The Si$\#$1 detector had 48 readout channels which consisted of the first 32 single strips readout and the remaining 32 strips which were read out pairwise by the remaining 16 channels. All 64 strips of Si$\#$2 were read out individually. Each germanium detector had 32 readout channels. The MPR16 and MSCF16 used for the strip readout are 16 channel preamplifiers and shaping amplifiers, respectively. A single channel preamplifier MPR1 was used to set bias voltage and read out the rear side of the detectors. The 32 channel MADC32 is a typical peak sensing ADC for energy measurement.

A customised eight channel ISEG~\cite{iseg} high voltage module EMS 82x1p-FSHV has been employed for providing the bias for the detectors. The first four channels with 500 V maximum output and the remaining four channels with 2000 V maximum output were dedicated to supply the bias voltages for the silicon and germanium detectors, respectively. The remote control via internet for the module integrated in an MPod crate has been realised with the Tcl/Tk programming language on a linux system. 

 A VME based DAQ system with common EMS framework~\cite{ems2,ems3} for COSY experiments has been established for the detector laboratory tests and the commissioning experiment at COSY.   All of the relevant hardware including a VME crate, peak sensing ADCs, a scaler module as well as a computer have been integrated into the data acquisition system. An online monitoring program dedicated to the commissioning experiment has been developed.

\section{Laboratory tests}
\label{sec:labtest}

 \begin{figure}
\begin{center}
\resizebox{0.48\textwidth}{!}{
  \includegraphics{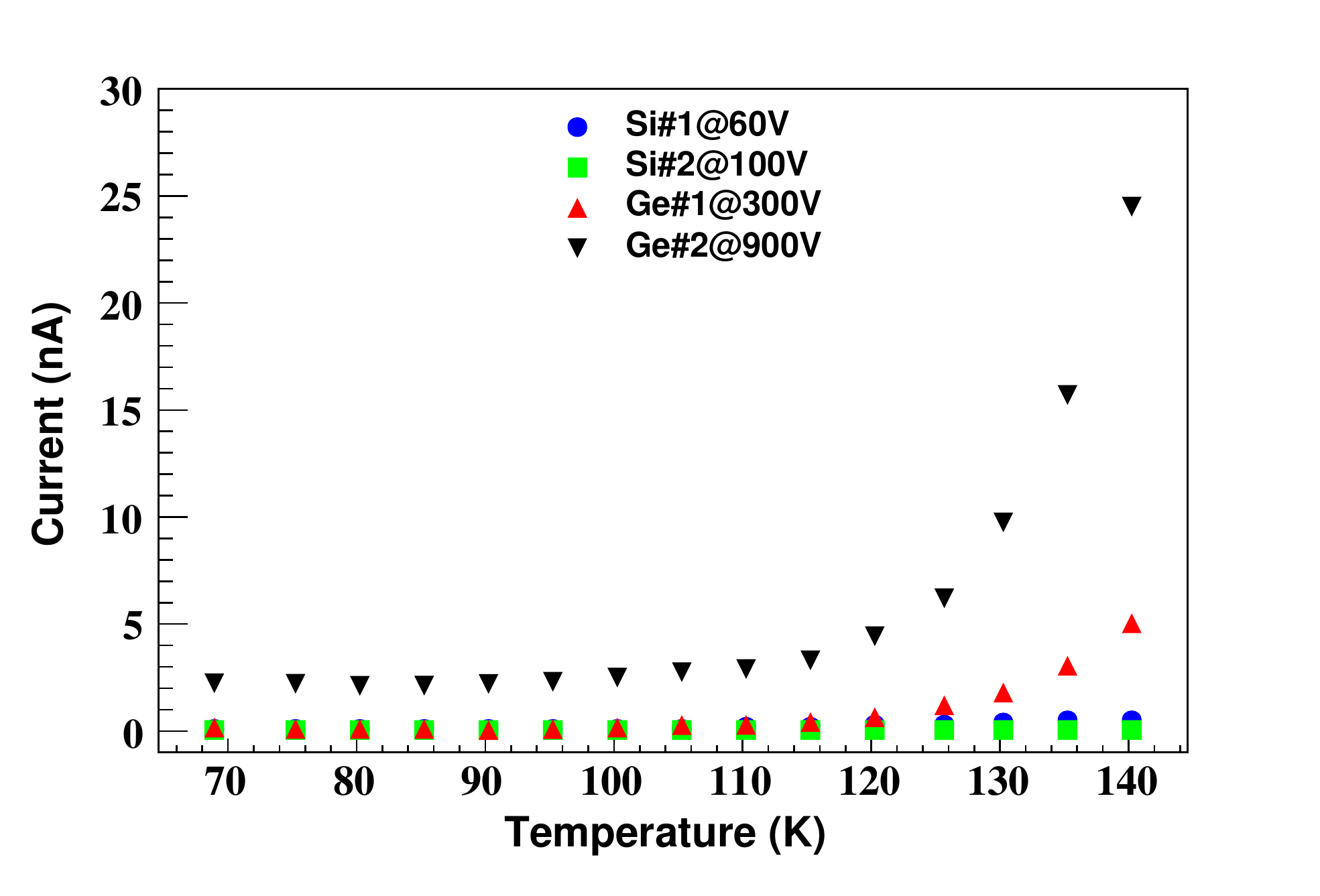}
}
\end{center}
\caption{Detector current versus operating temperature.}
\label{fig:8}   
\end{figure}

Prior to the final assembly of the detector system, single detector checks have been performed with a test chamber including a liquid nitrogen dewar. The tests implemented for the silicon detectors consisted of a measurement of current versus voltage as well as response to an alpha radioactive source. The behaviour of the silicon detectors at 300~K and 105~K was consistent with expectations. The leakage current of the silicon detectors significantly decreased when they were cooled to 105~K. The current of the silicon detectors versus operating temperature in the range of 70--140~K is shown in Fig.~8. Germanium detectors are in general operated at temperature of 77--100~K by using liquid nitrogen. The response of the germanium detectors has been examined at 100~K by using an alpha radioactive source. After inspecting all single detectors, the final assembly was made. To obtain precise energy measurement on the strips and rear side of the detectors, special efforts have been put into the working temperature and the energy calibration of the detectors.

\subsection{Working temperature of the germanium detectors} 
\label{sec:workingtemp}

 \begin{figure}
\begin{center}
\resizebox{0.48\textwidth}{!}{
  \includegraphics{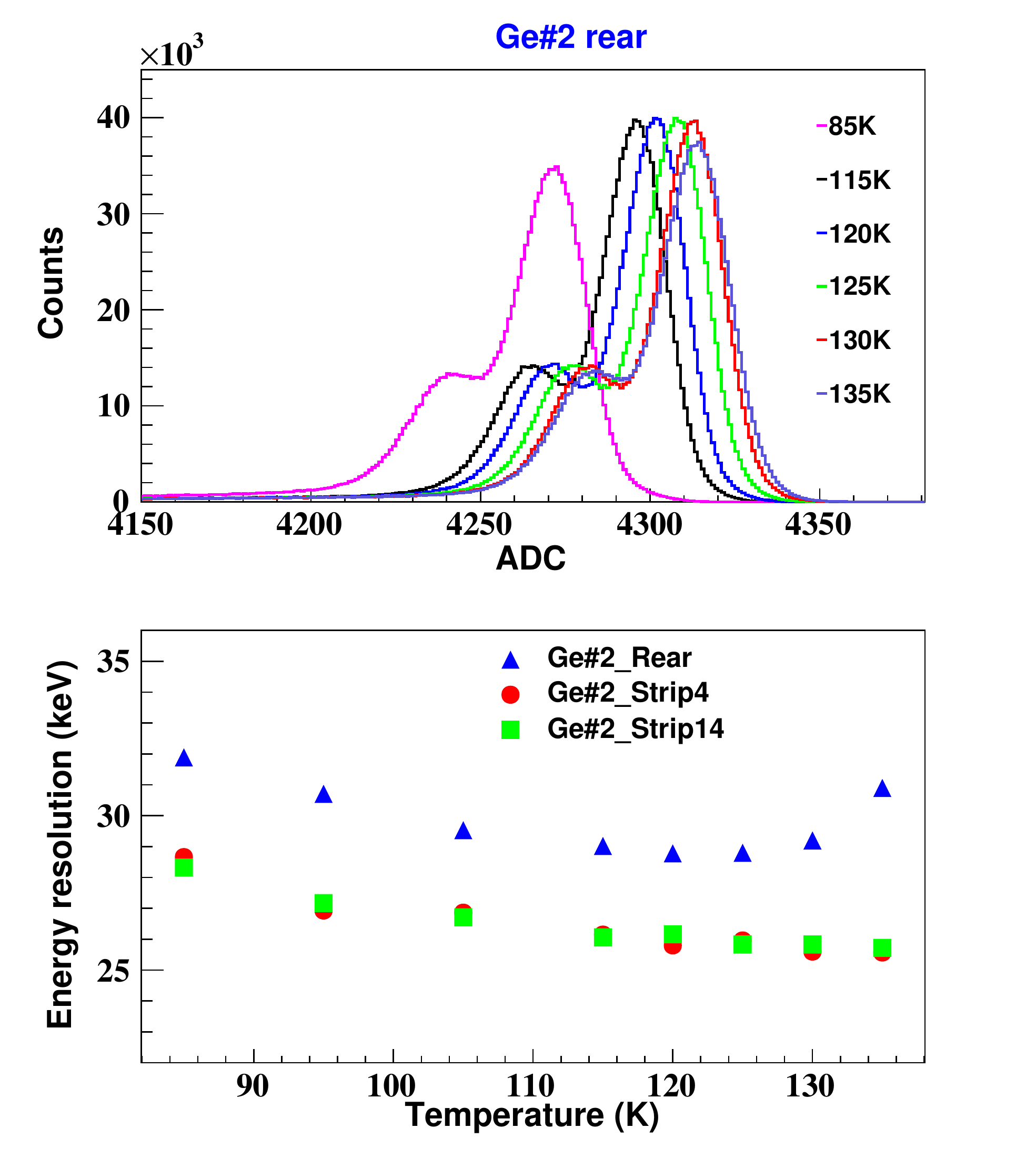}
}
\end{center}
\caption{ADC energy spectra with a $^{244}$Cm alpha source on the rear side (upper) and the energy resolution for 2 strips as well as the rear side (lower) of the Ge$\#$2 (11~mm thick) detector at various working temperature.}
\label{fig:9}      
\end{figure}

One of the laboratory tests was to adjust the working temperature of the detector system in order to achieve the best energy resolution of the germanium detectors. Using a $^{244}$Cm alpha radioactive source, the energy resolution of the germanium detectors as a function of the working temperature has been measured in steps of 5~K. It was found that the germanium detectors have better performance at higher temperatures. The energy resolution of the strips has improved while the signal amplitude increased. However, the leakage current of the germanium detectors increased rapidly above about 120~K, as shown in Fig.~8. As a consequence, the energy resolution on the detector rear side deteriorated above about 130~K. The ADC spectra of the alpha energy on the rear side of the Ge$\#$2 detector at various working temperatures is displayed in the upper plot of Fig.~9. The lower plot of Fig.~9 shows the energy resolution for two strips and the rear side. It is clearly seen that the amplitude increases and the energy resolution improves not only on the strips but also on the rear side with increasing temperature until about 125~K. Those tests also showed that the silicon detectors had higher signal amplitude when the working temperature increased from 85~K to 135~K. However, no further improvement on the energy resolution of the strips could be found. Thus, 125~K was adopted as the working temperature for the recoil detector system. 

With these operation conditions, energy resolutions better than 20~keV and 30~keV (FWHM) for the silicon and germanium strips have been achieved, respectively. The performance of both types of detectors fulfils the design specifications.

\subsection{Energy calibration} 
\label{sec:calibration}

The energy calibration of the detectors was based on the known decay energies of radioactive sources and the pulse height linearity of the electronics. The silicon detectors have excellent energy resolution and an ultra-thin nominal window layer ($<$0.1~${\mu}$m), therefore, the energy calibration of the silicon detectors has been determined by using the 2 known alpha energies of 5.763~MeV and 5.805~MeV from a $^{244}$Cm source. 

\begin{figure}
\begin{center}
\resizebox{0.48\textwidth}{!}{
  \includegraphics{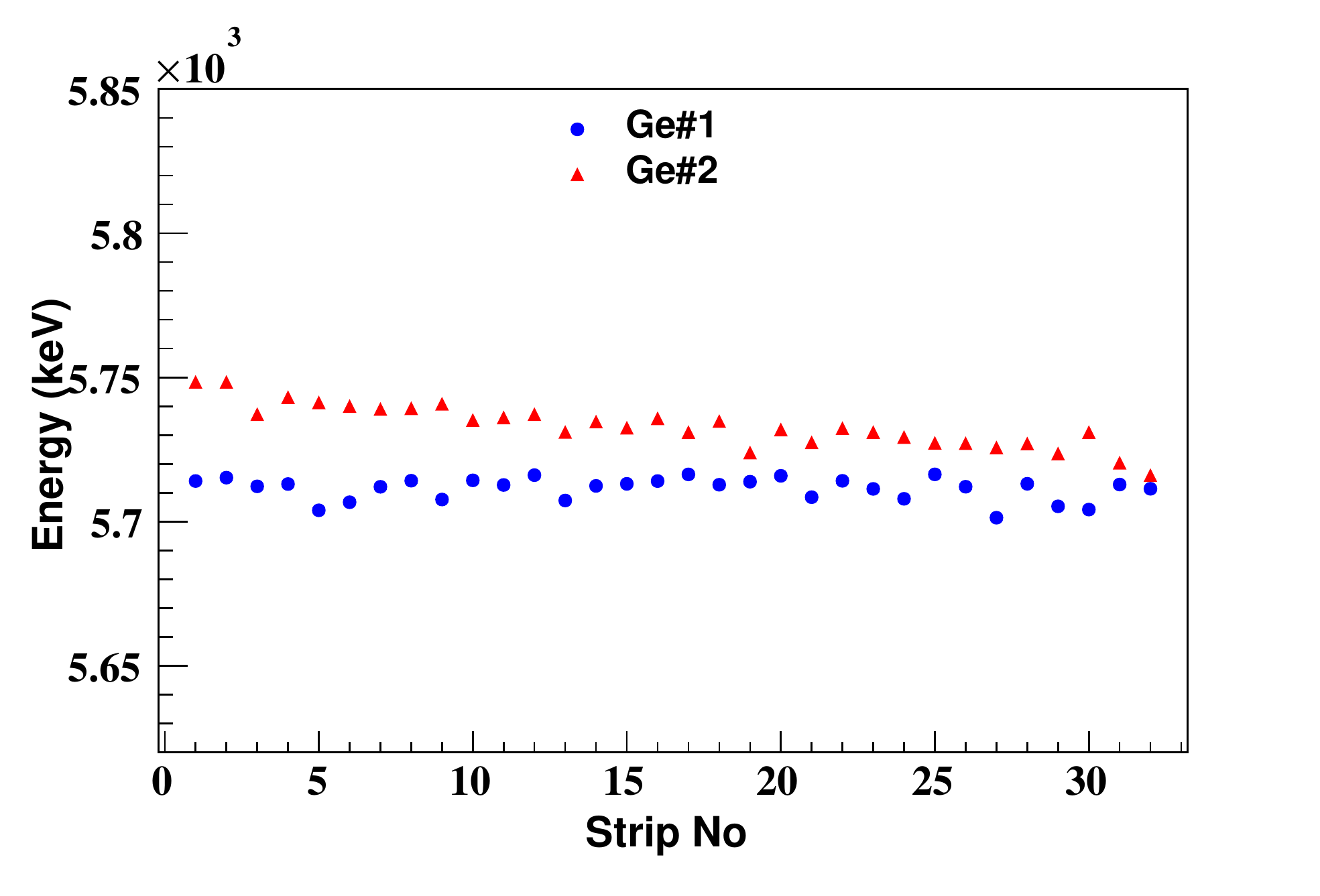}
}
\end{center}
\caption{Effective alpha energy deposited in each strip of the germanium detectors.}
\label{fig:10}    
\end{figure}

The germanium detectors have a nominal dead layout of about 1~$\mu$m on the strips, which results in a worse energy resolution of the germanium strips than the silicon strips. As a consequence the method mentioned above was not used to calibrate the germanium detectors. Alternatively, the germanium detector was calibrated by determining the effective energy that was deposited by alpha particle in the strips. The effective energy is the "remaining" energy of the alpha particles after passing through the dead layer of the detectors. We rely on the measured energies of the gamma rays from the radioactive sources of $^{137}$Cs with 662~keV and $^{60}$Co with 1.173~MeV and 1.332~MeV to be equal to the nominal energies since gamma rays have little sensitivity to the dead layer of the detectors. With the benefit of the excellent linearity of the electronics, the effective energy was determined by extrapolating the measured gamma energies to the measured amplitude of alpha particles from the source $^{244}$Cm. The measurements indicated that the effective dead layers of Ge$\#$1 and Ge$\#$2 are about 0.7 $\mu$m and about 0.4 $\mu$m, respectively. Fig.~10 shows the effective energy of alpha particle with nominal energy of 5.805~MeV deposited in each strip of the germanium detectors. The effective energies on Ge$\#$1 and Ge$\#$2 converge at high strip numbers due to the effects of the increasing incident angles of alpha particles on Ge$\#$2. After additional tests with a different gain setting of the electronics it was found that the precision of the energy calibration for the silicon and germanium detectors was better than 1$\%$.

\section{Beam tests}
\label{sec:beamtest}
 
The main goals of the tests with proton beams were to prove the validity of the detector concept and to study the detector response to the recoil protons, in particular for the following features: the minimum measurable proton energy, the precision of the energy measurement and the background suppression. The recoil detector was installed for a beam test at the existing hydrogen cluster target station of the ANKE experiment~\cite{anke} inside of the COSY ring.

\begin{figure*}
\begin{center}
\resizebox{0.98\textwidth}{!}{
\includegraphics{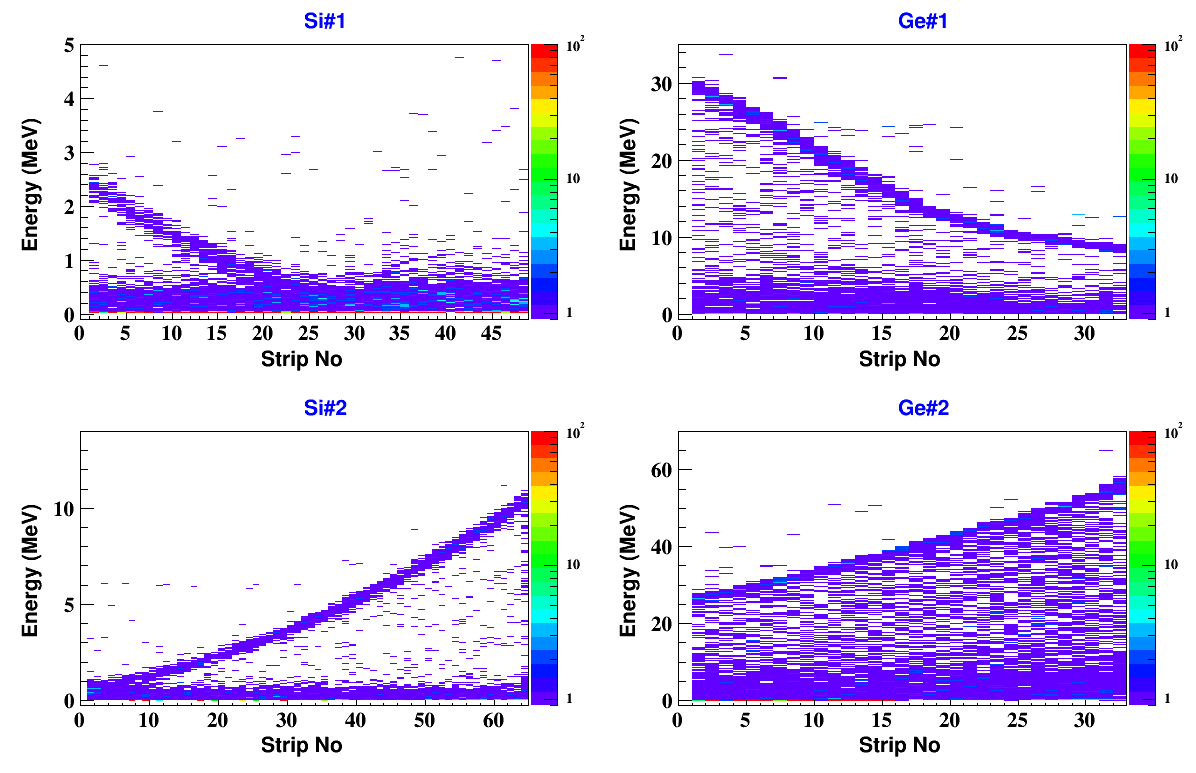}
}
\end{center}
\caption{Raw energy spectra measured by all 4 detectors.}
\label{fig:11}     
\end{figure*}

\subsection{Hydrogen cluster target and beam}
\label{sec:2}
 
The existing hydrogen cluster-jet target at the ANKE target station in the COSY ring was used for the commissioning experiment. 

\begin{itemize}
   \item {\bf Target specification} \\
In order to be consistent with the design requirements of the KOALA experiment, the cluster beam thickness along the beam axis should be less than 1 mm. The width of the cluster beam on the plane normal to the beam axis was about 9 mm~\cite{Khoukaz}. The desired cluster beam dimension was achieved with a customised collimator~\cite{Taeschner}. Another important feature of the target is the areal density, which was estimated to be $10^{14}$ atoms/cm$^2$ in this specific configuration. 
   \item {\bf Maximum recoil angle} \\
   The maximum recoil angle is the essential limit for the measurement of the $t$-distribution. Fortunately, the existing target chamber at ANKE allowed a maximum recoil angle of 13.6$^{\circ}$, which just meets the desired angular range for 3.2~GeV/c beam momentum. 
\end{itemize}

During the recoil detector commissioning, the COSY ring was operated with about 10$^{10}$ protons in the ring and a 5~mm beam diameter at the target position. A beam cycle with a flat top duration of 5 minutes was used for data taking.

\subsection{Data taking}
\label{sec:datataking}
 
Data of the proton-proton elastic scattering at the beam momenta of 1.7, 2.5, 2.8 and 3.2~GeV/c have been taken in two separate measurement weeks. The recoil detector system worked properly during the whole beam time. The raw energy spectra measured by the 4 detectors at a beam momentum of 3.2~GeV/c are presented in Fig.~11. The bands of energy deposition in the various strips corresponding to elastic scattering was clearly observed. It can also be seen that the fraction of counts under the bands on the strips of the germanium detectors increased. This was dominated by events where more than one strip fired due to the higher energy and larger recoil angle. 

\subsection{Analysis and results}
\label{sec:analysis}

The data analysis presented here was used to evaluate the data quality and to answer key questions which are essential to validate the proposed KOALA experiment. All following results are based on the data sample at a beam momentum of 3.2~GeV/c.

\subsubsection{Clustering algorithm}
\label{sec:clustering}

\begin{figure}
\begin{center}
\resizebox{0.48\textwidth}{!}{
  \includegraphics{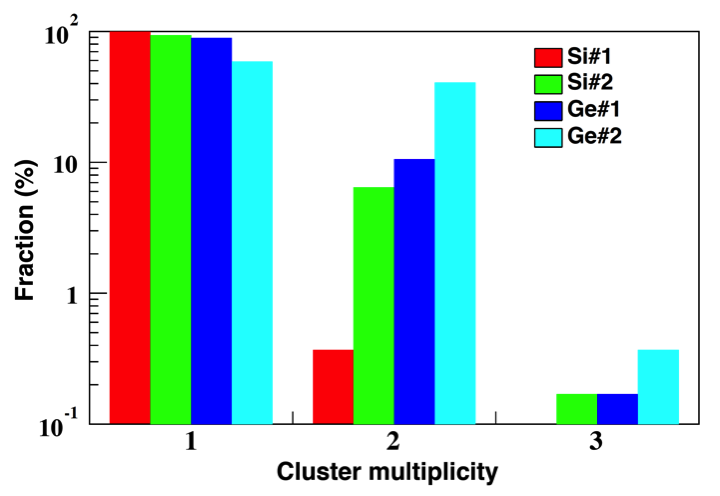}
}
\end{center}
\caption{The multiplicity distribution of clusters.}
\label{fig:12}      
\end{figure}

 An energy clustering algorithm has been implemented in order to reconstruct the proton energy deposited in more than one detector strip. An energy cluster consists of all relevant neighbouring strips in which the deposited energy is above a specific, strip dependent, threshold. The total energy of each event has been reconstructed based on the energy of the cluster. The hit distribution for all clusters for each detector is plotted in Fig.~12. As expected the strip multiplicity of incident particles on the germanium detectors is higher than on the silicon detectors due to the higher energy and the larger recoil angle. Most of the detectors had average multiplicities just above 1 except for Ge$\#$2 which had a large fraction of events with multiplicity 2. There were very few events with multiplicity 3, or higher.

After having performed the clustering, a preliminary energy spectrum over the full range of measurable recoil angles has been plotted in the lower part of Fig.~13. In contrast to the energy spectrum before clustering shown in the upper part of Fig.~13, it is demonstrated that the energies of events with multiplicity greater than 1 have been well reconstructed. The blank pattern on the lower right  corner of the plot after clustering is the consequence of the variable threshold for the clustering algorithm.

\begin{figure}
\begin{center}
\resizebox{0.48\textwidth}{!}{
 \includegraphics{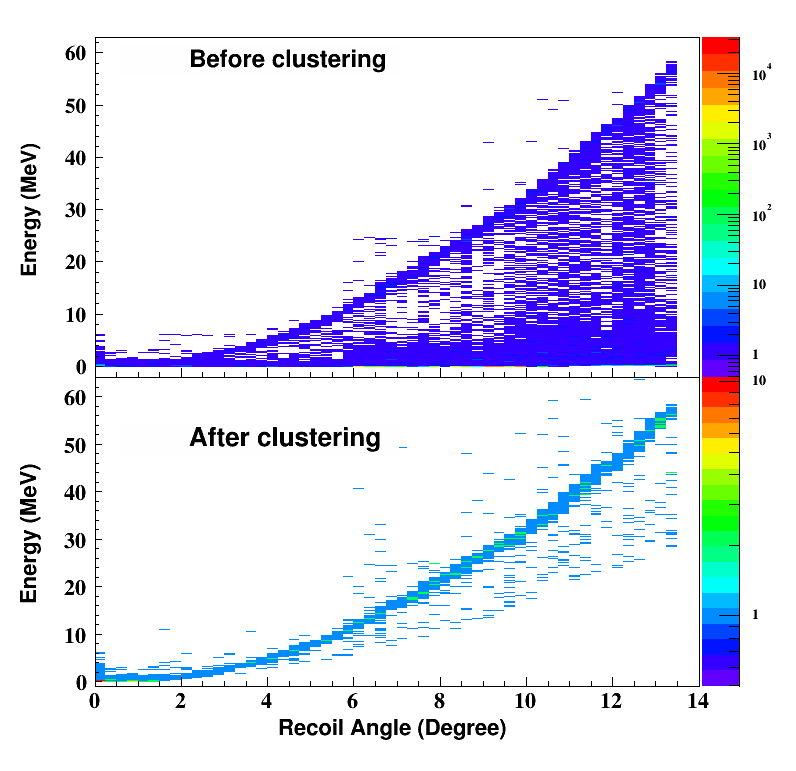}
}
\end{center}
\caption{The energy spectrum before and after clustering.}
\label{fig:13}       
\end{figure}

\subsubsection{Preliminary energy curve}
\label{sec:Ecurve}
In order to assess the precision of the measured energy of the recoil protons, a preliminary energy curve as a function of recoil angle is shown in the upper plot of Fig.~14. The measured energy in red is almost overlapping the calculated ideal energy in blue. The  lower picture of Fig.~14 shows the ratio of the measured energy to the ideal energy. The small discrepancy up to 2$\%$ between the spectra is considered to be due to several small effects, such as detector alignment and beam target position, which will be corrected in further analysis. Taking the calibration precision into account, a precision of about 1$\%$ of the measured energies of the recoil protons can be expected after correcting the small effects mentioned above.

\begin{figure}
\begin{center}
\resizebox{0.48\textwidth}{!}{
  \includegraphics{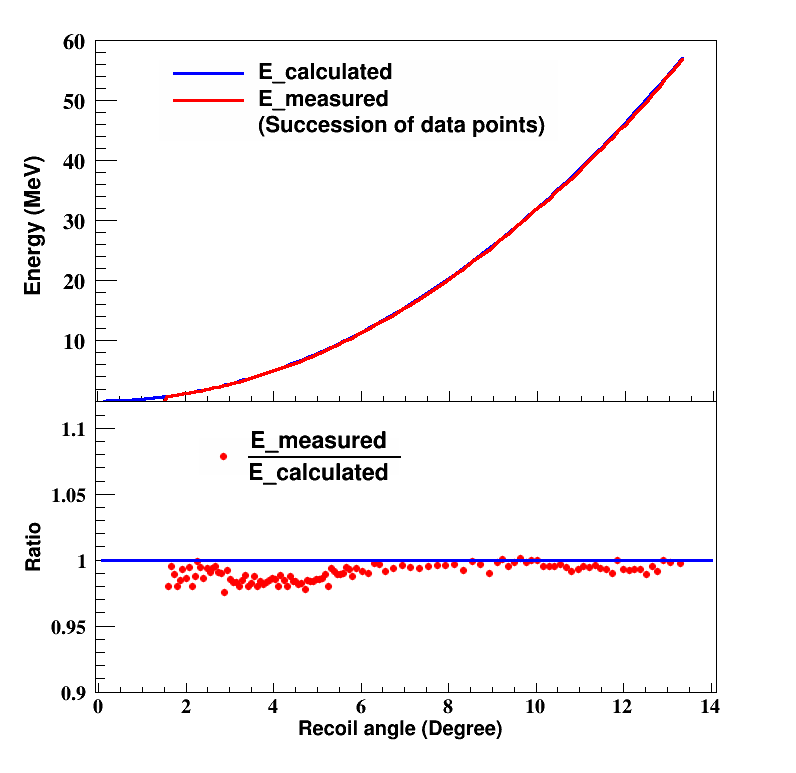}
}
\end{center}
\caption{(upper part) The blue and red (a succession of data points) solid line show the ideal energy curve and measured energy of recoil protons as a function of recoil angle, respectively. The ratio between the measured energy in a given strip to its corresponding ideal energy is presented in the lower part.}
\label{fig:14}     
\end{figure}

\subsubsection{Detector energy threshold}
\label{sec:6}
The detector energy threshold is the minimum energy of the recoil protons for which the elastically scattered protons can be identified. It appears as the background signal rate becomes comparable or even higher than the signal in the small recoil angle region. The criteria to evaluate the energy threshold is to judge whether the proton signal is separable from the background. Fig.~15 shows a plot of energy (after clustering) versus strip number measured by the first silicon detector at the smallest recoil angle as well as two single strip energy spectra. The two arrows on the energy plot of Si$\#$1 point to the strips No.~4 and No.~20. The  lower left histogram of Fig.~15 shows a clear recoil proton energy distribution on strip No.~4, which is completely separated from the background. The narrow width of the energy spectrum was dominated by the small strip width as well as the thin target. The background has been suppressed by the clustering algorithm.  The lower right corner of Fig.~15 depicts that the proton's energy measured by strip No.~20 partially overlaps the background. Therefore, the clustering algorithm was not able to easily separate the signal from the background. It indicates that it will be difficult to separate the interesting proton signal from the background when the recoil proton energy is lower than 600~keV. A coincidence with the forward measurement should enable a lower detector energy threshold. This can be expected when the full setup of the KOALA experiment at HESR is available.

\begin{figure}
\begin{center}
\resizebox{0.48\textwidth}{!}{
  \includegraphics{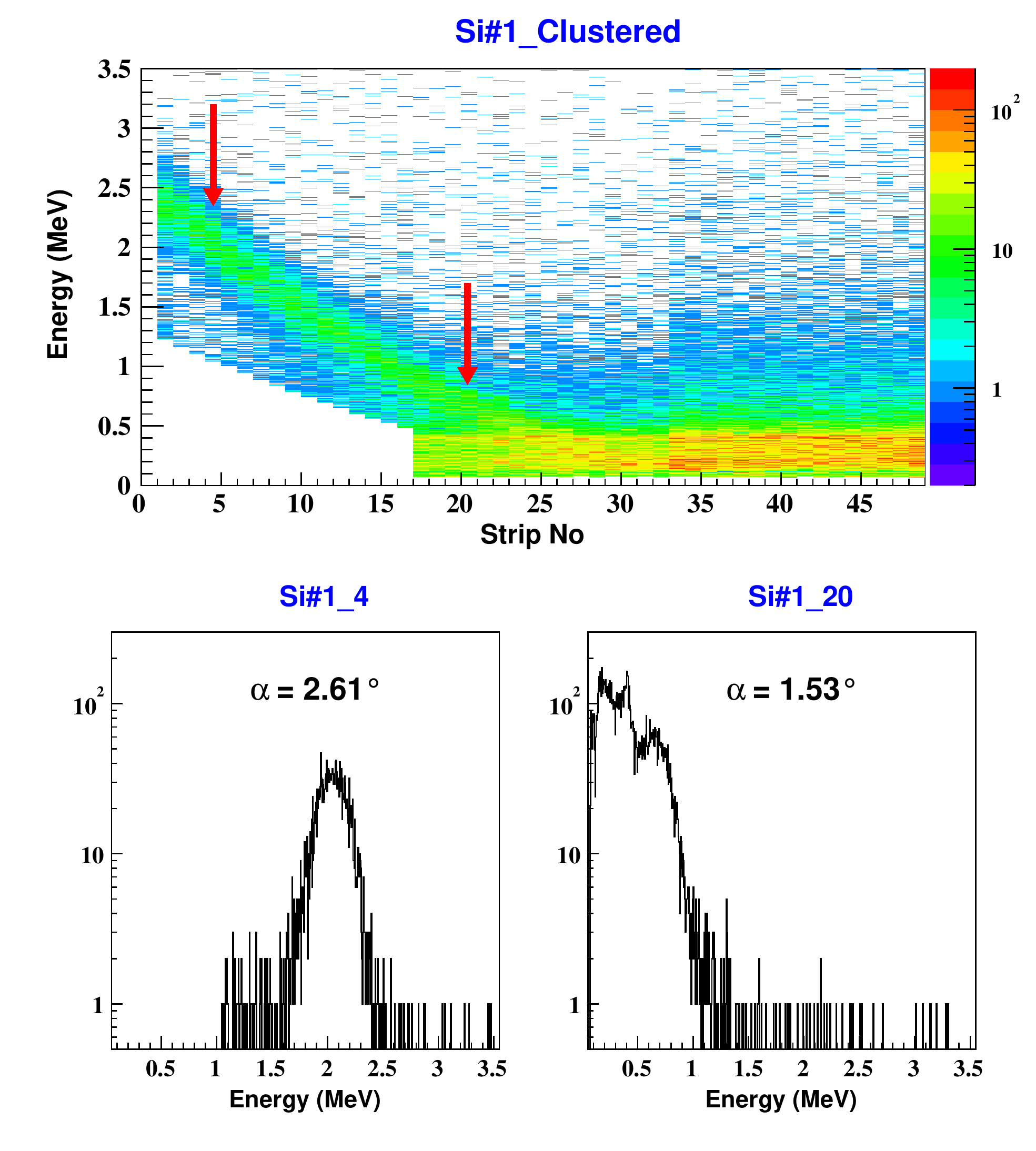}
}
\end{center}
\caption{ (upper) Histogram of proton energy versus strip number of the first silicon detector, which is in the small recoil angle region. The lower histograms present the energy histograms of strip No.~4 (lower left) and No.~20 (lower right). These strips are marked by the red arrows in the upper frame.}
\label{fig:15}       
\end{figure}

\section{Conclusion and outlook}
\label{sec:conclusion}
 
In order to test the method proposed for the KOALA experiment at HESR, a recoil detector to measure the recoil protons of antiproton-proton elastic scattering at scattering angles close to 90$^{\circ}$ has been designed and built. Laboratory tests have demonstrated that the performance of the recoil detector meets the experimental requirements. Excellent energy resolution better than 20 keV and 30 keV (FWHM) of the silicon strip detectors and germanium strip detectors have been achieved, respectively. Proton-proton elastic scattering has been measured at COSY. The detector system was working properly. No observation of a variance of the leakage current as well as the energy resolution of the detectors was found during the beam test.  The measured energy of the recoil protons is in good agreement with the expected energy. The preliminary result shows a discrepancy between measured and ideal energy smaller than 2$\%$. The recoil detector has also shown a good background suppression capability. Even without requiring a coincidence with the forward scattered proton, an energy threshold of 600 keV for the recoil protons was achieved. All these results validated the concept of the KOALA experiment at HESR. 
 
Further data analysis is going on to reconstruct the $t$-distribution. In order to achieve the best precision on the $t$-distribution, the detector alignment as well as simulation studies on the beam-target overlap position are being performed. The luminosity and elastic scattering parameters will be determined when the $t$-distribution is available. A coincidence between the recoil detector and the forward measurement will lower the energy threshold, and thus the minimum $t$ value measurable. This will be realised by the full setup of the KOALA experiment at HESR.

\label{sec:acknowledgement}
\begin{acknowledgement}
We are grateful to the COSY crew who installed the device into the COSY ring and provided proton beams for tests. We owe the ANKE collaboration who allowed the installation of the recoil detector chamber at their target station for a beam test. Special thanks for S. Mikirtytchiants, who gave the excellent support to find the optimal beam-target overlap. Many thanks are dedicated to D. Chiladze for the Schottky measurements. This work was performed for part of the PhD thesis of Qiang Hu. \end{acknowledgement}

\end{document}